\documentclass[a4paper,11pt]{article}
\pdfoutput=1 

\usepackage[utf8]{inputenc}
\usepackage{upgreek}
\usepackage{jinstpub}

\newcommand{\krthree}{${}^\mathrm{83m}$Kr}
\newcommand{\1}[1]{\ensuremath{\mathrm{#1}}}
\newcommand{\mus}{\1{\upmu{}s}}
\newcommand{\degree}{$^\circ$}

\title{The XeBRA platform for liquid xenon\\time projection chamber development}

\author{Daniel~Baur,}
\author[1]{Alexander~Bismark,\note{Now at Physik-Institut, Universität Zürich, 8057 Zürich, Switzerland}}
\author{Adam~Brown,}
\author{Julia~Dierle,}
\author{Horst~Fischer,}
\author{Robin~Glade-Beucke,}
\author{Jaron~Grigat,}
\author[2]{Basho~Kaminsky,\note{Previously at Albert Einstein Center for Fundamental Physics, Universität Bern, 3012 Bern, Switzerland}}
\author{Fabian~Kuger,}
\author{Sebastian~Lindemann,}
\author{Darryl~Masson,}
\author{Patrick~Meinhardt,}
\author{Mariana~Rajado~Silva,}
\author{Marc~Schumann,}
\author{Florian~Tönnies and}
\author{Francesco~Toschi}

\affiliation{Physikalisches Institut, Universität Freiburg, 79104 Freiburg, Germany}

\emailAdd{adam.brown@physik.uni-freiburg.de}
\emailAdd{fabian.kuger@physik.uni-freiburg.de}

\abstract{%
XeBRA is a flexible cryogenic platform designed to perform research and development for liquid xenon detectors searching for rare events.
Its extra-large outer cryostat makes it possible to install a wide variety of detector designs.
We present the system, including its cryogenic, gas handling, data acquisition and slow control subsystems.
Two dual phase time projection chambers with sensitive masses at the 1~kg scale have so far been operated in XeBRA.
Using data from these, we determine the field-dependence of the electron drift velocity in liquid xenon.
We also measure the relative charge and light yields for 41.5~keV energy deposits from \krthree{} with electric drift fields between 50~V/cm and 677~V/cm.
}


\begin{document}
\maketitle
\flushbottom

\section{Introduction}
\label{sec:intro}

Low-background time projection chambers (TPCs) filled with the liquid noble gas xenon~\cite{XENONExpt, XenTERResults, LUXExpt, LZFirstResults, PandaXIIResults, PandaXFirstResults} provide the strongest limits on the cross-section of spin-independent interactions of WIMP dark matter with masses from a few GeV up to several TeV.
Experiments currently taking data will be roughly an order of magnitude more sensitive after an exposure of around five years~\cite{XenTSensitivity,LZSensitivity}.
In the future, multi-ton experiments such as DARWIN~\cite{DARWINExpt} plan to probe the entire experimentally accessible range of cross-sections for a large range of WIMP masses~\cite{LXeWhitepaper}, while other experiments such as nEXO~\cite{nEXOSensitivity} will use the same technology to search for neutrinoless double-beta decay.

An interaction in the liquid xenon target of a dual-phase TPC produces scintillation light, ionisation electrons and heat.
The first two of these can be measured to reconstruct properties of the interaction.
The scintillation light is directly detected, usually by photomultipler tubes (PMTs), and is commonly called the S1 signal.
Electrons are drifted towards the liquid surface at the top of the TPC using an electric field, known as the drift field.
Here, a second, stronger electric extraction field pulls them into the xenon gas above the liquid target where they are accelerated.
These two electric fields are created by a set of three electrodes, referred to as the cathode, gate electrode and anode.
The drift field is between the cathode and gate electrode, while the extraction field is between the gate and anode.
Their collisions with gaseous xenon atoms produce a second light signal, proportional to the number of electrons, which is also detected using the PMTs.
This is known as the S2 signal.
The energy deposited in an interaction can be determined from the strength of these two signals.
Their relative sizes give insight into the type of interaction involved.
The distribution of the S2 light on the PMTs can be used to reconstruct the lateral $(x, y)$ position of the event, while the time delay between the two signals indicates its depth $z$.

As TPC-based experiments grow larger and aim for sensitivity to lower cross-sections, the background level must also be reduced.
This leads to the exploration of a variety of new technologies, some examples being radial designs with a single central anode wire~\cite{RadialTPCProposal, RadialTPC}, alternative light sensors such as silicon photomultipliers~\cite{SiPM_TPC}, and hermetic TPCs~\cite{HermeticTPC, HermeticQuartzTPC, HermeticUCSD}.
Before being used in a large experiment, these new concepts must first be developed and tested on a small scale.
The XeBRA platform at the University of Freiburg, described in section~\ref{sec:xebraplatform}, provides a facility for such studies in its 250~l outer cryostat vessel, which insulates the cryogenic liquid xenon detector.
So far, kilogram-scale TPCs have been operated in an 11~l inner vessel, which can easily be enlarged to test larger detectors. They are presented in section~\ref{sec:tpcs}.

These prototype detectors also allow the detailed study of the liquid xenon physics involved in TPC operation.
For example, the velocity of electrons as they drift, which is an important input when optimising a detector's design.
Similarly, knowledge about the way the deposited energy is distributed between the light and charge signals is important for simulations~\cite{NEST}, required both for detector design and for understanding the data from existing liquid xenon TPCs.
Results from such measurements are presented in section~\ref{sec:results}.

\section{The XeBRA platform}
\label{sec:xebraplatform}

\begin{figure}
    \centering
    \includegraphics{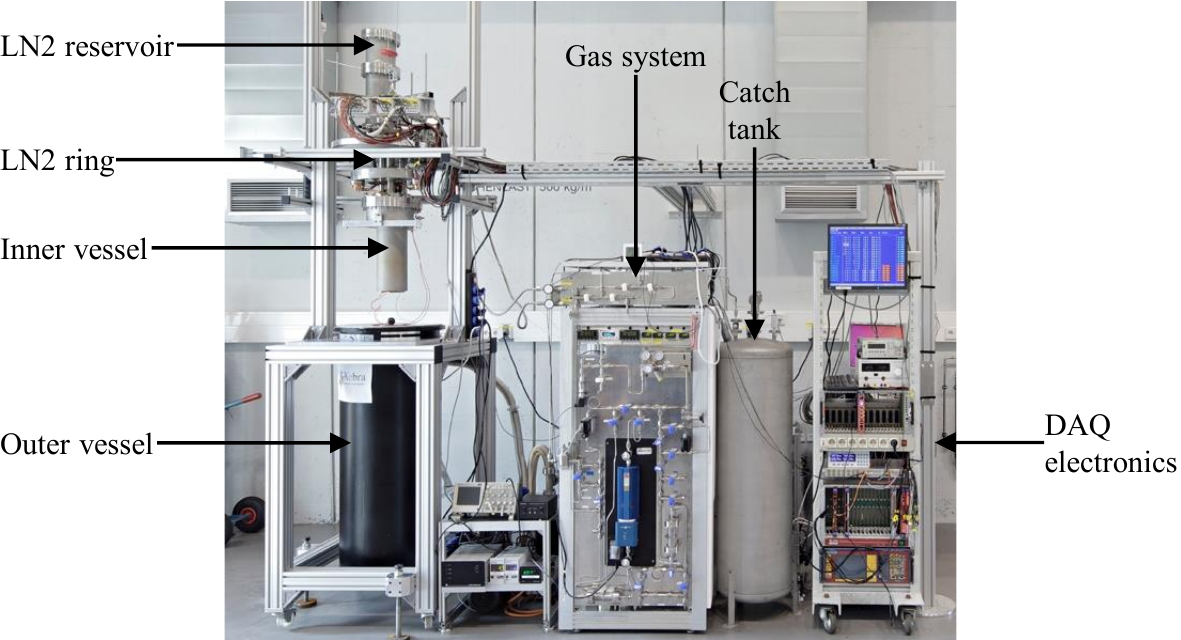}
    \caption{The XeBRA platform. On the left, the inner and outer vessels are both visible, and the spare room provided by the large outer vessel is apparent. The gas system is visible in the centre, the two storage bottles are hidden behind the aluminium panel. The catch tank can also be seen, and the 19-inch rack on the right houses data acquisition and slow control electronics.}
    \label{fig:setupphoto}
\end{figure}

The xenon-based research apparatus (XeBRA) allows testing of liquid xenon TPCs of the ten-kilogram scale.
Seen in figure~\ref{fig:setupphoto}, it consists of a double-walled cryostat, liquid-nitrogen-based cooling, and a xenon handling system.
A flexible slow-control system continuously monitors detector parameters, while PMT signals are acquired using a trigger-less data acquisition system.

\subsection{Cryogenics}

\begin{figure}
    \centering
    \includegraphics{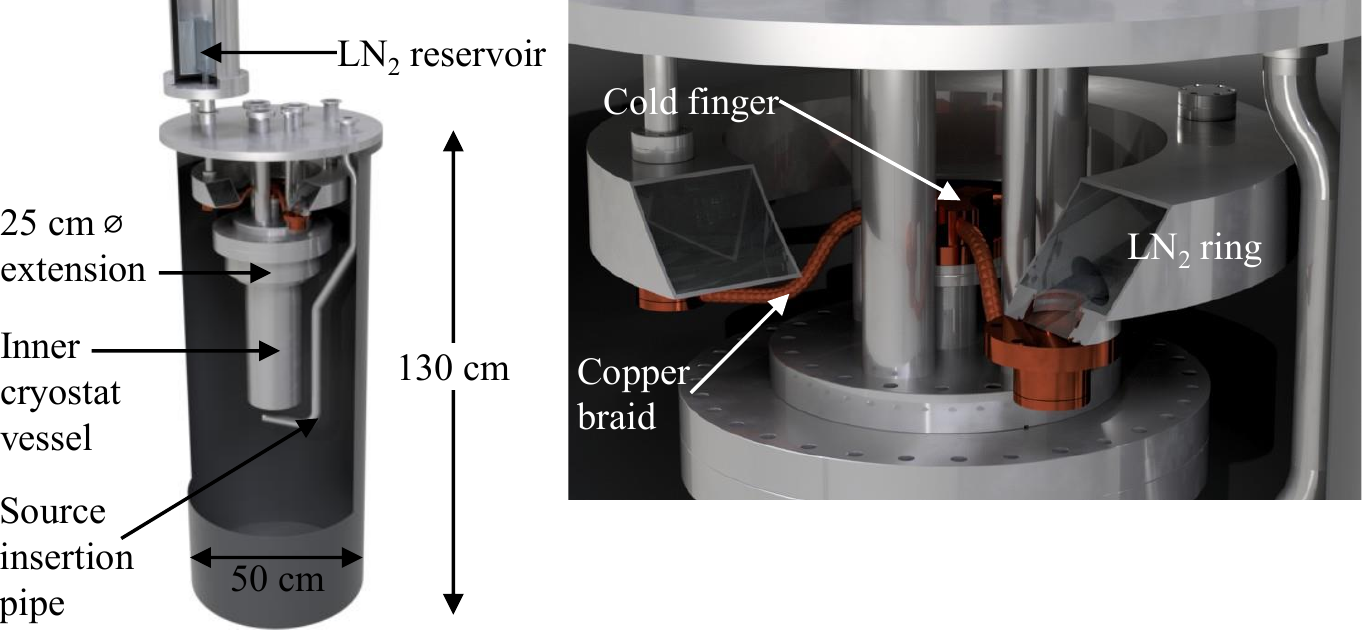}
    \caption{The XeBRA cryostat, with the zoom on the right showing additional detail. The outer vessel is cut open to show the inner vessel and the liquid nitrogen cooling ring with copper braids connecting it to the top of the cold head. Several vacuum tubes pass through both parts of the cryostat allowing cable and gas connections between the inner vessel and the laboratory outside. The source-insertion pipe, seen on the right, allows calibration sources to be positioned at any height along the side of, or underneath the TPC.}
    \label{fig:topflange}
\end{figure}

The outer vacuum vessel of XeBRA is a 50~cm diameter and 130~cm tall aluminium cylinder, sealed with an FKM O-ring.
These large dimensions allow flexibility for future tests, for which the inner vessel could easily be enlarged.
The inner cryostat vessel's top CF-150 flange is suspended from the top flange of the outer cryostat.
The current stainless-steel inner vessel is also cylindrical, with a diameter of 15~cm and height of 39~cm.
At the top, an 8~cm extension with 25~cm diameter provides additional space for cable management.
Both cryostat vessels and various ancillary components are illustrated in figure~\ref{fig:topflange}.
One 16~mm and three 40~mm diameter tubes pass through the outer vessel's upper flange and provide access to the inner vessel for cabling and xenon gas routing.
An additional 25~mm diameter tube passes down the side of the inner cryostat and underneath it, to allow radioactive sources to be positioned for calibration.
The inner vessel is wrapped in superinsulation consisting of multiple layers of metallised plastic foil to suppress radiative heat transfer.
The total heat intake of around 100~W is dominated by conduction along the tubes between the outer and inner flanges.

The platform is installed on a frame built from aluminium profiles.
To access the detector, the top flange of the outer vessel is first lifted using an electric winch, with the inner vessel still hanging from it.
After fixing the flange in its raised position, the inner vessel is accessed by opening its flange and lowering it back into the outer vessel using three hand-operated winches.

The system is cooled by liquid nitrogen, held in a hollow reservoir ring within the isolation vacuum, surrounding the four feed-through tubes.
Three copper braids connect the ring to a copper cold finger with direct contact to the xenon gas inside the inner vessel.
During operation, a Cryo-Con 22C PID controller controls a 50~W heater to stabilise the cold finger temperature.
The liquid nitrogen ring is fed from a 4~l open-flask reservoir on the top of the main flange.
This in turn is refilled periodically from an external pressurised dewar, with a solenoid valve allowing automated refilling.
A Teragon LC2 controller triggers the refill when the liquid nitrogen level drops below a temperature sensor within the reservoir.
A capacitive level meter, which is independent of the refilling control, also monitors the liquid nitrogen level using the slow control system described below.
All the critical electronics, including the refilling control and monitoring, is connected to an uninterruptible power supply, such that the cooling can tolerate short power outages.
This enables stable and safe operation of the platform over several weeks.

\begin{figure}
    \centering
    \includegraphics[trim={5.77cm 2.83cm 13.45cm 0.79cm},clip]{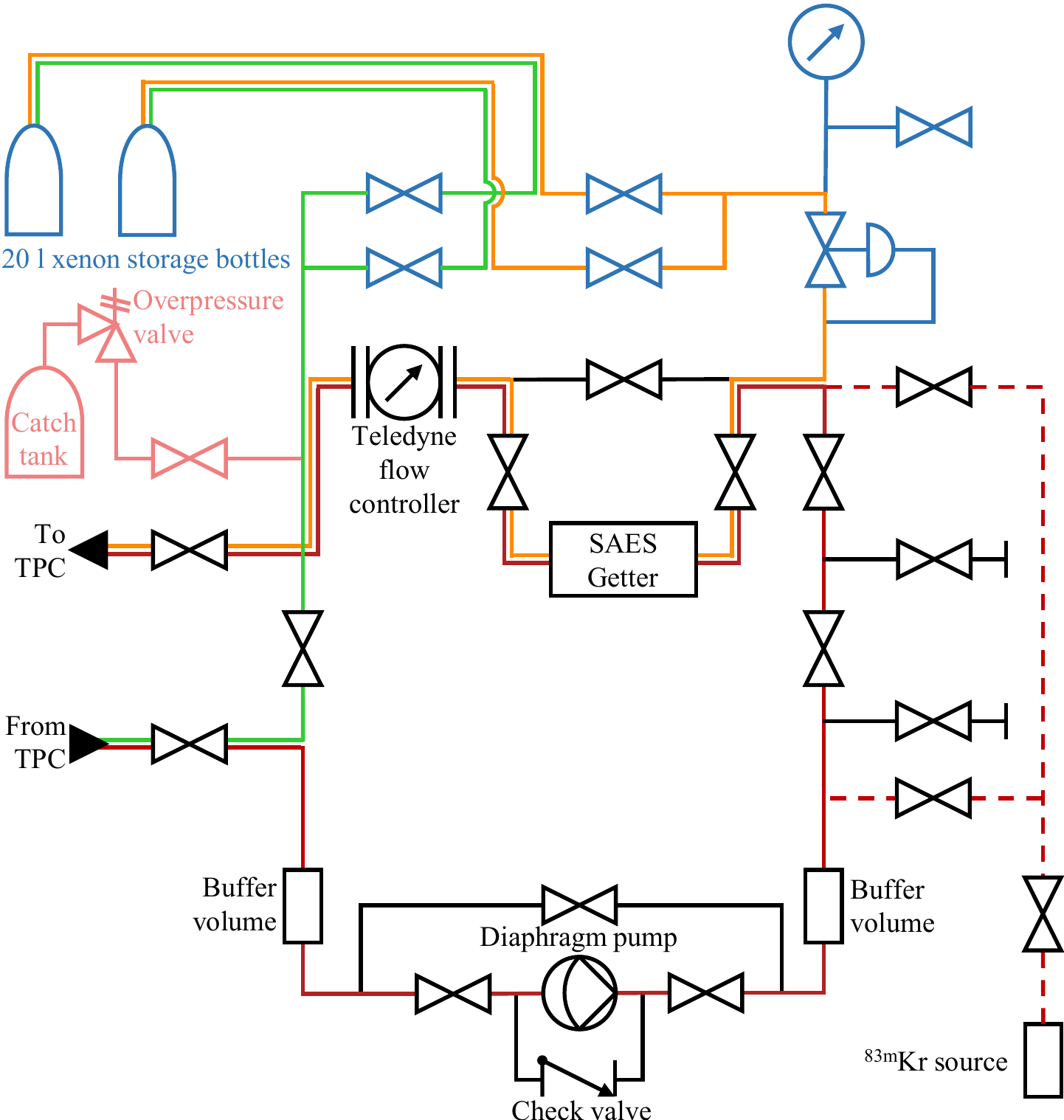}
    \caption{Piping and instrumentation diagram of the xenon-handling system. The flow path used during stable operation is indicated in red. The orange flow path is used when filling xenon into the cryostat, and the green path when recovering xenon from the cryostat. Components which are part of the (high-pressure) xenon storage system are coloured blue, and the emergency pressure relief system and catch tank is pink. The diaphragm pump for recirculation, depicted at the bottom, can be bypassed if necessary and is connected in parallel with a check valve to avoid dangerous differential pressures which could damage the diaphragm. The four valves on the right allow auxiliary systems to be connected; here a \krthree{} source is attached to two allowing xenon flow to be directed past the source (dashed red line). These valves can also be used to attach a pump in order to evacuate the gas system. By only opening one of the valves, the \krthree{} injection rate can be reduced, thanks to the longer diffusion path.}
    \label{fig:pid}
\end{figure}

Around 10~kg of xenon is sufficient to fill the inner cryostat vessel once a typical detector such as those described here has been installed.
A xenon-handling system enables the filling, purification, recovery and storage of xenon from the cryostat.
The system is constructed using 1/4-inch stainless steel pipes and 1/4-inch VCR components, with fittings and valves supplied by Swagelok.
It is schematically illustrated in figure~\ref{fig:pid}, where the flow paths during filling, stable operation and recovery of xenon are indicated.
During stable operation, a \hbox{KNF N022} membrane pump extracts xenon from the cryostat and drives it through the purification loop, designed for flows up to around 5~slpm (standard litres per minute).
A Teledyne Hastings HFC-302 flow controller is used to restrict the flow.

A SAES MonoTorr PS3-MT3 heated zirconium getter removes electronegative impurities before the xenon is returned to the cryostat, and can also be used when initially filling.
Such impurities can trap electrons while they drift, reducing the size of the S2s observed with the mean electron absorption time being referred to as the electron lifetime.
For the TPCs discussed in this work, clean xenon gas is directed into an enclosed volume around the cold finger where it is liquefied.
A pipe injects the clean liquid xenon directly into the TPC at around the height of the cathode.
By directly liquefying and injecting clean xenon, the time required to reach an acceptable purity in the detector after filling it with liquid xenon is reduced by a factor~20, compared to the case where a mixture of newly-cleaned and existing gas is liquefied and not injected into the TPC.
Typically, the purity is sufficient to achieve electron lifetimes of around 80~\mus{} or better after a few days.
Xenon is recovered from the TPC by freezing the xenon gas in one of the storage bottles using liquid nitrogen and cryogenically pumping from the cryostat.

A custom built, multiple-tube-in-tube heat exchanger is fitted into one of the 40~mm pipes connecting the inner volume with the main top flange.
The liquid xenon leaving the cryostat passes through six thin-walled copper pipes of 3~mm diameter.
Here it evaporates, absorbing heat from the incoming gas, which cools and condenses on the copper pipes.
Up to 65\%~of the energy needed to evaporate the xenon removed from the system is harvested from the input flow.
This efficiency decreases for xenon flows above 2~slpm, due to the shorter transit time of the gas through the heat exchanger.
A higher efficiency would be difficult to achieve with the existing space constraints, and this is sufficient for the current uses of the platform.
To ensure xenon purity, the heat exchanger is constructed without glue or brazing.
Instead, the design exploits the different thermal contraction coefficients of its materials to seal by cryofitting.

An evacuated 115~l tank is connected to the gas system via an overpressure relief valve.
This safety system prevents dangerous pressures being reached if, for example, cooling power is lost, while avoiding xenon loss and exposure to air.

\subsection{Slow control, alarm system and long term operation}

Operational parameters, such as pressures, temperatures, voltages and currents, filling levels of liquid nitrogen and liquid xenon, xenon flow rates and heat load, are monitored by a versatile and lightweight slow control system based on python and custom-developed for the system~\cite{JGrigatThesis, DobermanPaper, Doberman}.
This lightweight system makes it straightforward to incorporate a wide variety of sensors and allows some processing of the recorded data.

Alarms are distributed by email and SMS when parameters cross pre-defined thresholds.
As an example, if the liquid nitrogen supply runs out or the refilling fails, an alarm is sent as soon as the level in the tank drops below its usual range.
This gives operators around two hours to fix the problem before the pressure and temperature begin to rise in the cryostat.
Additionally, all data are permanently stored in a time-series database using InfluxDB~\cite{influx}.
They can be visualised live online, both in text form and graphically via the open-source visualisation software Grafana~\cite{grafana}.
Some aspects of the platform can also be remotely controlled, notably the temperature set point of the cold finger.
This enables a faster response in the event of an unexpected change.
Thanks to the slow control system, the platform can be operated largely unsupervised.

\begin{figure}
    \centering
    \includegraphics{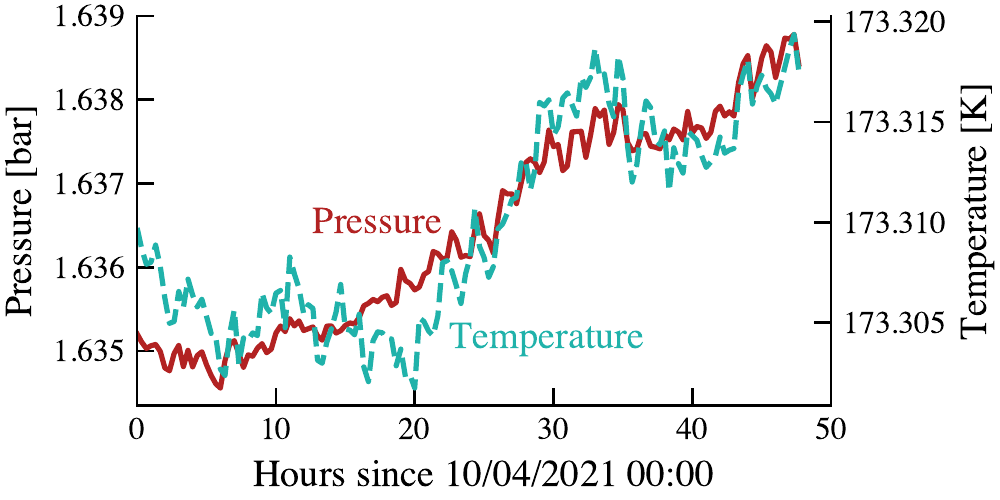}
    \caption{%
    Pressure (red) and temperature (blue) variation in the XeBRA cryostat during an example two-day period.
    The residual long-time variations seen here are on the level of 0.01\% (temperature) and 0.2\% (pressure) over the two days, which has a negligible effect on the detector performance. They are the result of small environmental fluctuations such as the temperature in the laboratory environment. 
    }
    \label{fig:slowcontrol}
\end{figure}

Between 2018 and 2021 XeBRA was operated for several runs, the longest spanning more than three months.
The critical parameters were sufficiently stable over this period so as not to influence detector operation.
Residual fluctuations result from external temperature changes, due to varying weather or ventilation efficiency.
These induce long-term temperature variations in the inner cryostat vessel of around 0.02~K, which in turn result in pressure changes of around 0.005~bar, as seen in figure~\ref{fig:slowcontrol}.

\subsection{Data acquisition}

TPC prototypes operated in the XeBRA platform are typically equipped with PMTs to detect the S1 and S2 scintillation liqht signals.
The system for the readout of PMT signals is a scaled-down version of the trigger-less XENON1T data acquisition system (DAQ)~\cite{Xe1TDAQ}.
The PMTs are independently powered using a CAEN A1535DN high-voltage supply. 30~AWG Kapton-insulated wires supplied by Accu-Glass provide the connection between SHV sockets and the PMTs.
To reduce the number of wires, the neutral return from the PMTs is connected internally and carried to the power supply over a single wire.
RG196A/U coaxial cables with 50 $\upOmega$ impedance carry PMT signals to the DAQ.
Both the coaxial cables and Kapton wires pass through a potted vacuum feed-through produced by RHSeals, which has a leak rate better than 10$^{-8}$~\1{mbar\:l/s}.

Signals are amplified by a factor of~10 using a Phillips 776 amplifier, and then recorded by a CAEN V1724 flash digitiser with 100~MHz sampling rate and 14~bit resolution.
The digitiser firmware allows independent readout of all channels exceeding the self-trigger thresholds, without the need for a coincidence between PMTs to generate a global trigger.
The thresholds are set to a fraction of the expected amplitude of a single photoelectron, so that almost all PMT signals are digitised and saved for further analysis.
The lack of an additional software-based trigger is the sense in which the DAQ is trigger-less.
The digitiser is read out over an optical link using the redax readout software~\cite{redax}.
Processed data are constructed on a second server following the procedure discussed in section~\ref{sec:results}, and stored for later analysis.

\section{Dual-phase time projection chambers}
\label{sec:tpcs}

\begin{figure}
    \centering
    \includegraphics{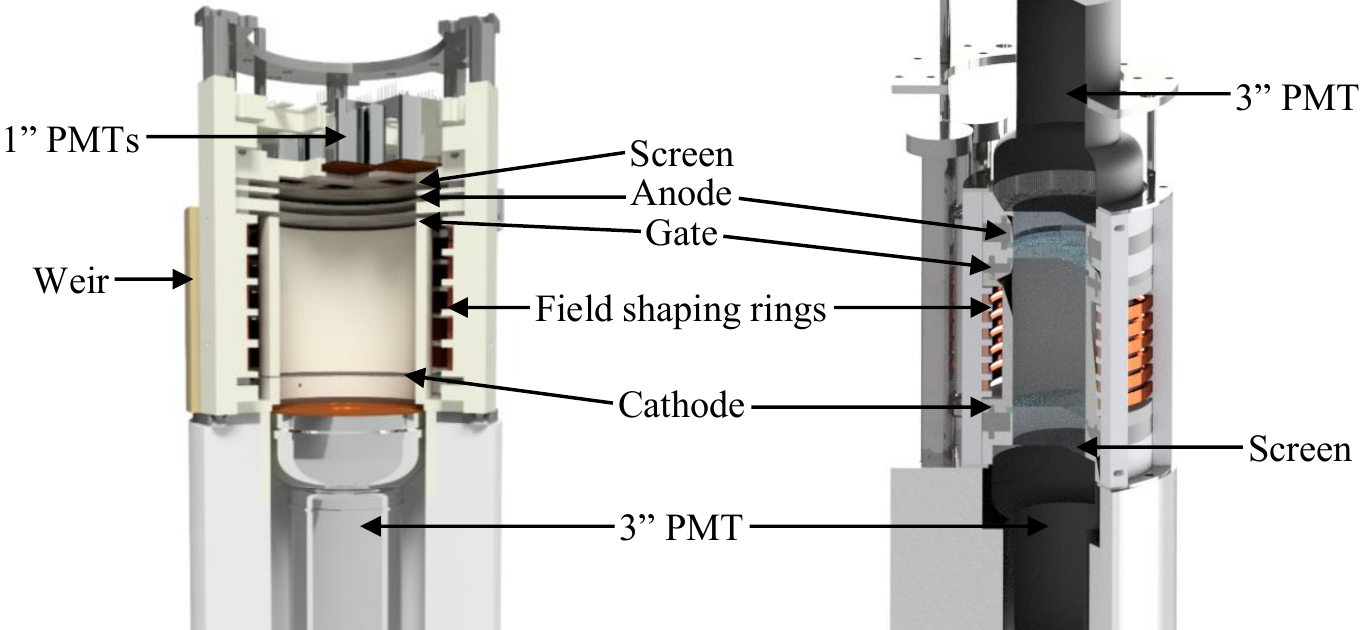}
    \caption{Cut-away view of the baseline dual-phase TPC (left) and hermetic TPC (right)~\cite{HermeticTPC} used in XeBRA. The sensitive region is between the cathode and gate electrodes. PMTs at the top and bottom record light signals following interactions.}
    \label{fig:tpcrender}
\end{figure}

In this section, we illustrate the flexibility of the XeBRA platform with two small-scale time projection chambers.
These serve as examples of the kind of detector which can be operated in XeBRA, and provide the data used for the results in section~\ref{sec:results}.
The ``baseline'' dual phase TPC serves as a benchmark against which future technologies can be tested in XeBRA, and has a similar design to the detectors used by current liquid-xenon dark matter experiments.
The second TPC discussed is the ``hermetic'' TPC, which aims to reduce the radon concentration in its sensitive region and is described in further detail and with results in~\cite{HermeticTPC}.
Both detectors are shown in figure~\ref{fig:tpcrender}.

\subsection{Baseline TPC}
\label{sec:baselinetpc}

The sensitive region of the TPC consists of a cylinder enclosed by a PTFE reflector with a 70 mm internal diameter, and defined at the bottom by a cathode and 70 mm above by a gate electrode.
Both these dimensions and all others given in this section refer to room temperature.
The cathode and gate electrode are used to create the vertical electric field which drifts ionisation electrons towards the top of the TPC.
The anode, 5~mm above the gate, creates the electric field required to extract electrons into the gaseous xenon and induce secondary scintillation.
A screening electrode, 5~mm above the anode, shields the top PMTs from electric fields and improves the field homogeneity.
All electrodes are hexagonal grids, etched from 150~\1{\upmu{}m} stainless steel sheets.
These are spot-welded to 3~mm thick stainless steel rings, and electrically isolated from each other by PTFE spacers.
The grid elements have a 150~\1{\upmu{}m} width and 3~mm pitch, resulting in an optical transparency of 90\%.
Five copper field-shaping rings outside the TPC wall ensure that the drift field within the sensitive region is sufficiently uniform.

The cathode is biased using a CAEN N1470 high-voltage supply, and a potential divider defines the voltage of each field-shaping ring.
The CAEN N1470 is also used to individually positively bias the anode and negatively bias the top screening electrode.
In typical operation conditions, a voltage of 3.5~kV between the cathode and gate electrode results in a drift field of 470~V/cm.
Between the gate and anode, a voltage difference of 4.5~kV provides a secondary scintillation multiplication field (in the gaseous xenon) of 11~kV/cm. 

During operation, the liquid-gas interface is located centrally between the gate electrode and anode, 2.5~mm from each.
A weir sets the liquid level; it consists of a hollow reservoir which is closed at the bottom and open at the top. When the liquid xenon level reaches the top of the weir it flows down into it.
A pipe from the bottom of the weir extracts xenon into the purification loop.
The xenon level in the TPC is monitored by three short capacitive level meters, with an active range of 8~mm, and one long level meter, covering the full TPC height. 
Four PT100 sensors monitor the temperature at the bottom of the inner cryostat vessel and at three positions along the TPC, giving insight into the temperature gradient while cooling and filling.

A single Hamamatsu R11410-21 PMT covers 84\% of the bottom of the TPC volume with its 64~mm diameter photocathode.
It is placed inside a solid aluminium filler to minimise the volume to be filled with liquid xenon, with a PTFE cylinder insulating the negatively biased PMT from the aluminium.
An array of seven Hamamatsu R8520 PMTs, each with a square 20.5~mm by 20.5~mm photocathode, provides 76\% geometrical coverage at the top.
Each of the eight PMTs can be biased individually to independently adjust their gains.
Two blue LEDs within the cryostat can be driven with a pulse generator to provide controlled light pulses and corresponding trigger signals.
These are needed to calibrate the PMTs' gains, using the statistical method of~\cite{ModelIndpendentGain}.

\begin{figure}
    \includegraphics{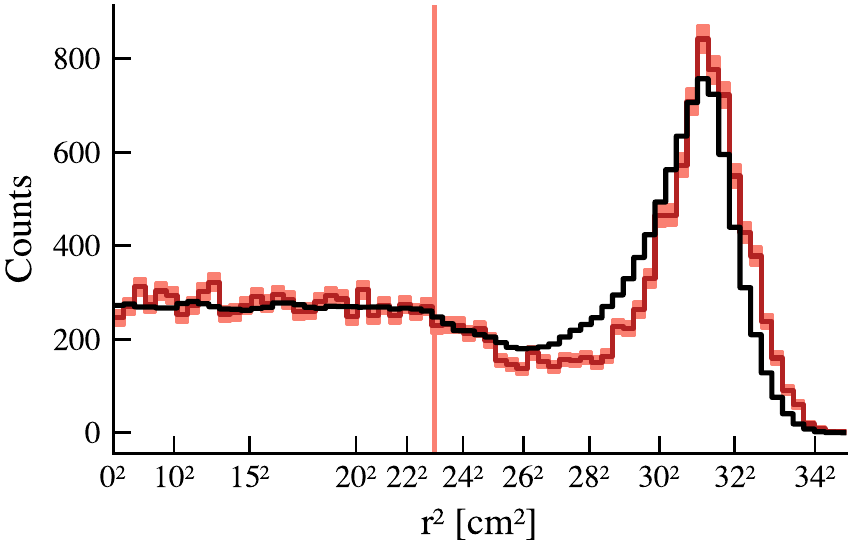}
    \hfill
    \includegraphics{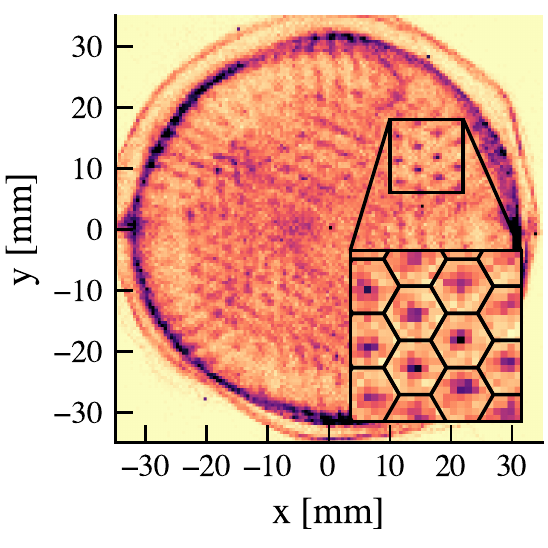}
    \centering
    \caption{%
        Left: reconstructed radius from the neural network of measured \krthree{} events (red) and of simulated events (black) in the baseline TPC.
        Right: reconstructed $(x, y)$ position of the same events, with the inset showing a zoomed region.
        At radii smaller than 23~mm (vertical pink line) the reconstruction preserves the true homogeneous distribution of the events.
        Events close to the edge of the TPC tend to be reconstructed on a circle at a radius of roughly 31~mm.
        As seen in the $(x, y)$ plot this circle is not perfectly round due to the hexagonal arrangement of PMTs in the top array.
        The inset is overlaid with a true-to-scale hexagonal grid to guide the eye.
        Electrons are funnelled through the centre of the gate electrode's mesh openings, resulting in a hexagonal pattern of reconstructed positions.%
    }
    \label{fig:posrec}
\end{figure}

The optical properties of the baseline TPC have been studied in detail using GEANT4~\cite{GEANT4}.
These simulations are used to train a Keras-based artificial neutral network~\cite{keras} to reconstruct interactions' horizontal $(x,y)$ positions~\cite{ABismarkThesis}. 
Two 64-node hidden layers in a dense layout have proven to provide the most accurate reconstruction.
Simulated S2s with 10$^4$ photons are reconstructed with a mean error of less than 0.8~mm for events within the inner $r \leq 23~\1{mm}$ cylinder.
The resolution is difficult to determine for real data where the true position is unknown, but is sufficient to resolve the hexagonal pattern of the gate electrode, as seen in figure~\ref{fig:posrec}.
Due to position reconstruction artefacts near the outside of the TPC, events with a reconstructed radius larger than 23~mm are not used for the results in this work.

\subsection{Hermetic TPC}

The second TPC used to collect data for the results presented in this work is the so-called hermetic TPC, also seen in figure~\ref{fig:tpcrender}.
This was developed as a proof of principle for mechanically separating the xenon inside the sensitive region from that outside the TPC.
Radon is a significant source of background for dark-matter searches and is emanated by all materials, especially outside the TPC where there is a larger surface area in contact with the xenon~\cite{Xe1TRadon}.
A hermetic TPC can therefore reduce the background rate within the sensitive region and improve a detector's sensitivity to dark matter.
This concept and its implementation as well as implications for future dark matter searches are discussed in~\cite{HermeticTPC}.
For the purposes of this work, the hermetic TPC is a second dual-phase TPC, with a generally similar design to the baseline TPC.
The main differences are outlined below.

The sensitive region of the hermetic TPC is also cylindrical, with a diameter of 56~mm and a distance of 75~mm between the cathode and gate electrode.
The anode is placed 5~mm above the gate; there is no screening electrode.
As in the baseline TPC, these electrodes are all hexagonal meshes etched from a 150~\1{\upmu{}m} stainless steel sheet with a pitch of 3~mm.

The liquid level is also defined using a weir.
The level in the sensitive region is monitored by a single capacitive level meter.
Due to the hermetic design, both the weir and the level meter are part of the hermetic inner volume, connected through pipes fixed to the anode and cathode support rings.
There is only a single 3" Hamamatsu R11410-21 PMT at both the top and bottom of the TPC.
As a result, no $(x,y)$ position reconstruction is possible.
However, the geometrical coverage of the top sensor is higher than that of the array deployed in the baseline TPC. 

\section{Liquid xenon response and electron drift properties}
\label{sec:results}

We make use of data collected by both the dual-phase baseline TPC and the hermetic TPC to measure certain properties of signal production in liquid xenon.
We provide field-dependent measurements of the electron drift velocity and present measurements of the charge and light yields at 41.5~keV, again accounting for their field-dependence.

All the results in this section make use of data collected with \krthree{} injected into the TPC.
 \krthree{} is produced during the decay of $^{83}$Rb and carried into the TPC with the recirculating xenon as shown in figure~\ref{fig:pid}.
Since the \krthree{} is expected to distribute homogeneously in the liquid xenon in the sensitive region, the positions of interactions should be uniformly distributed throughout the possible range.
The two-stage decay consisting of a 32.1~keV step followed by a 9.4~keV step, with an intermediate half-life of 156.8~ns~\cite{NuclearData83}, makes it possible to reliably tag events from \krthree{} decays based on the two S1 signals.
Only events containing two identified S1 signals with at least 20~PE (photoelectrons detected) and an S2 signal with at least 500~PE are used for the analysis presented here.
Due to their longer duration, the S2 signals from the two decays almost always overlap and are therefore reconstructed as a single S2 signal.
Therefore, we only present results for the combined energy of 41.5~keV.

Our data processor is based on that used by XENONnT~\cite{XenTERResults, strax}.
First, individual PMT signals separated by less than 150~ns are merged into one peak.
These are classified as S1 or S2 signals based on their duration, which is defined as the time between 25\% and 75\% of the photoelectrons having been detected.
S1s require a duration of at most 150~ns while S2s must be longer than 200~ns.
Finally, close enough signals are merged to form events.
The maximum separation of signals within an event is large enough to pair S1s with S2s generated after electrons have drifted the full length of the TPC.
These arrive with a delay of around 40~\mus{} to 55~\mus{}, depending on the drift field.
Electrons drift in liquid xenon at a constant speed, the drift velocity.
The depth $z$ of an interaction is therefore proportional to the drift time, the time between the first S1 and the S2.
In the baseline TPC, the horizontal position $(x,y)$ is found from the S2~signal's distribution over the top PMT array using a neural network, as described in section~\ref{sec:baselinetpc}.
Only events with a reconstructed radius less than 23~mm are used for the analysis described here, to avoid artefacts such as field distortion which occur near the edge of the TPC.
No horizontal position information is available for the hermetic TPC, since it only has one top PMT.

These results also rely on the area of the S1 and S2 signals, defined as the time-integral of the voltage waveform.
They are determined by integrating the individual PMT signals and correcting for the gain.
The areas from all PMTs which contribute to a signal are summed, and corrected to account for position-dependent effects.
The S1 area is roughly twice as large at the bottom of the TPCs than at the top.
This is because many upward-travelling photons from interactions near the top are reflected from the liquid xenon surface.
They therefore have a long travel path, requiring them to cross three electrodes and potentially be reflected off the wall more times before being detected, increasing their chance of being absorbed.
The amount of S1 light detected is found to be largely independent of the transverse position $(x, y)$ in the baseline TPC, and in the hermetic TPC this information is not available, so only the $z$-dependence is corrected.

A similar correction for the S2 accounts for electrons being absorbed by electronegative impurities in the xenon as they drift.
Assuming a constant absorption probability, the fraction of electrons reaching the top of the TPC follows an exponential distribution, with a mean absorption time, or electron lifetime, of typically 50~\mus{} or longer.
This is comparable to the time electrons need to drift the full length of the TPC, which is between 40~\mus{} and 55~\mus{} depending on the field.
The parameters for both the S1 and S2 corrections are determined from \krthree{} calibration data.
Since the energy deposited from such decays is always the same, the amount of light and charge produced per decay is uniform throughout the constant-field drift region, making a study of position-dependent detector events straightforward.
The corrected areas, called cS1 and cS2, are used throughout the remainder of this work.

\subsection{Drift velocity}
\label{sec:driftvelocity}
By comparing the drift times of events occurring at the cathode and at the gate electrode we determine the drift velocity of electrons in liquid xenon $v_\mathrm{D} = \Delta z / (\Delta t_\text{cathode}-\Delta t_\text{gate})$ as a function of the drift field.
The distance between the gate and cathode $\Delta z$ is 69~mm in the baseline TPC and 74~mm in the hermetic TPC.
Both values account for the 1.4\% linear contraction of PTFE when being cooled from room temperature to about $-100$~\degree{}C ~\cite{PTFEExpansion}.

An event's drift time $\Delta t$ is the delay between the first S1 and S2.
The S2 time is defined as the point at which 50\% of the total area has been detected.
This compensates for any diffusion, which is equally likely to push electrons to shorter and longer drift times.
There is no such diffusion for the S1 signal, and the photon travel time in the TPC is small compared to the typical S1 duration of tens of nanoseconds.
Therefore, the shape of S1 signals is independent of the event position and the exact definition of the S1 time is less important: different definitions would introduce only a constant offset to both the gate electrode and cathode drift times which would cancel out in the determination of drift velocity.
However, due to the sharp rise and slower fall of S1 signals, the start time is least affected by statistical fluctuations in the time-distribution of photons being detected.
We therefore use the time that the first PMT triggers to define the S1 time.

We determine the drift time $\Delta t_\textrm{gate}$ of events occurring at the gate electrode (gate events) using their S2/S1 ratio. 
The strong extraction field above the gate electrode results in a higher charge yield and lower light yield for decays in that region than below the gate, leading to a higher S2/S1 ratio.
The extraction field ``leaks'' a certain distance into the drift region, resulting in a gradual transition of the field strength.
Since the leakage is not symmetrical around the gate electrode, it is difficult to precisely identify the drift time of events occuring at the height of the gate electrode
Instead, we define the gate drift time $\Delta t_\text{gate}$ as the position of the S2/S1 transition itself and correct the distance $\Delta z$ for the fact this is not identical to the position of the electrode..

\begin{figure}
    \centering
    \includegraphics{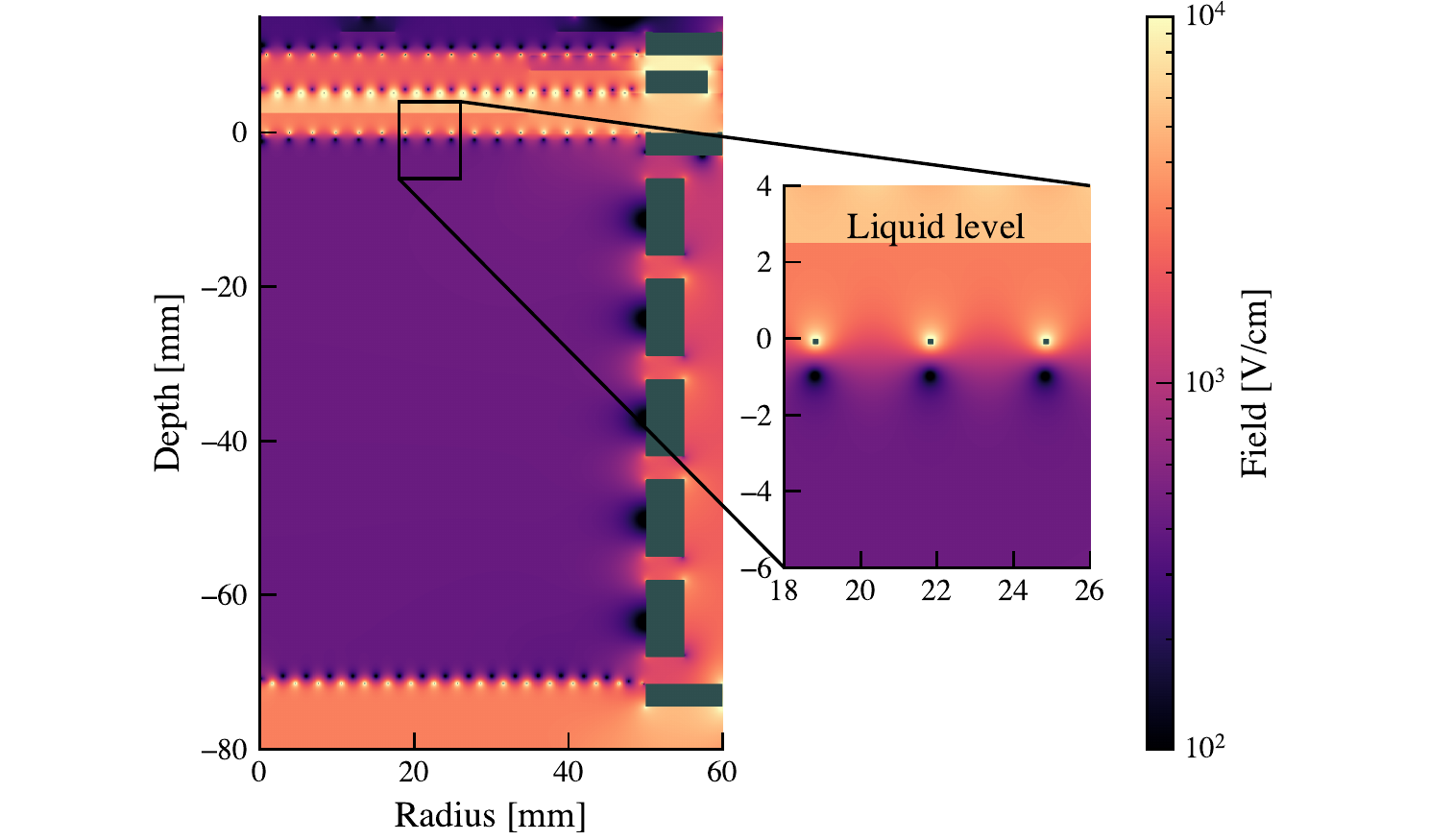}
    \caption{COMSOL simulation of the electric field in the baseline TPC (left) and zoom into the region around the gate electrode (right). Conducting elements are shown in grey, including the field shaping rings, electrode rings and electrode wires. The simulations are performed in two dimensions, assuming axial symmetry, meaning the hexagonal mesh electrodes are approximated by regularly spaced wires, visible in grey in the zoom. The gradual transition between the high and low field regions above and below the gate is clearly visible in the zoom. The field in the gaseous xenon is higher due to the lower dielectric constant. Other step changes, such as above the screening mesh at radii greater than 35~mm, follow the edges of PTFE elements.}
    \label{fig:comsol}
\end{figure}

\begin{figure}
    \centering
    \begin{minipage}[t]{0.47\textwidth}
        \centering
        \includegraphics{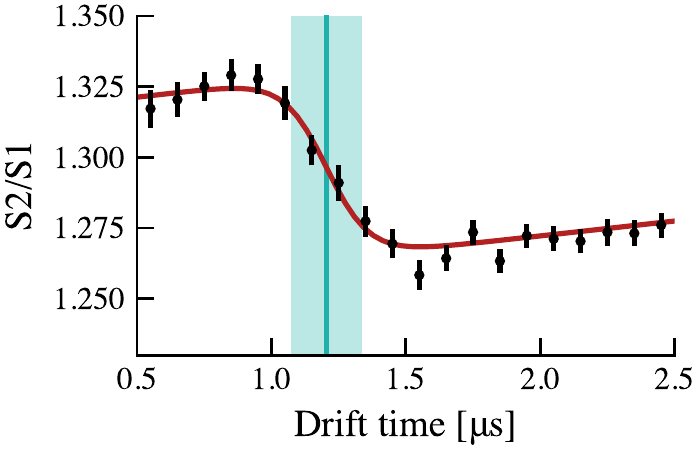}
        \caption{%
        Example of determining the drift time of electrons from the gate electrode in the hermetic TPC with a drift field of 595~V/cm.
        The mean $\text{S}2/\text{S}1$ ratio varies in the drift-time-dependent electric field as shown by the black data.
        It is fit with a Fermi function (dark red line), which is multiplied by a linear function representing the varying light detection efficiency.
        The blue shaded region indicates the 16th to 84th percentile range, used to determine the top of the drift region.
        The 50th percentile is indicated with a solid line.
        }
        \label{fig:gateposition}
    \end{minipage}
    \hfill
    \begin{minipage}[t]{0.47\textwidth}
        \centering
        \includegraphics{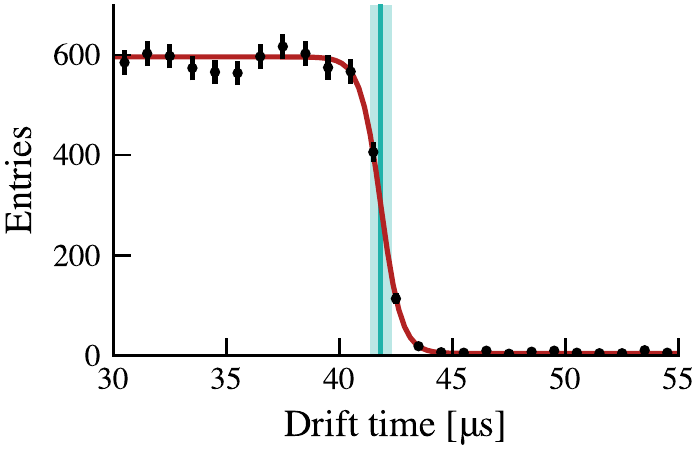}
        \caption{%
        Example showing how the cathode drift time is determined in the hermetic TPC with a drift field of 595~V/cm.
        The histogram (black points) shows the number of events within each drift-time bin with statistical uncertainties.
        A Fermi function is fit to the histogram (dark red line).
        Its $(50 \pm 25)$th percentile defines the drift time of cathode events (blue line and shaded region), in this case $(41.8 \pm 0.5)$~\mus{}.
        }
        \label{fig:cathodeposition}
    \end{minipage}
\end{figure}

For this, we simulate the electric field in the region around the gate electrode using COMSOL~\cite{comsol}, seen for the baseline TPC in figure~\ref{fig:comsol}.
The corresponding depth-dependent S2/S1 ratio can then be determined using NEST simulations~\cite{NEST} and can be described using a Fermi function.
The top of the electron drift region used for this study is defined as the 50th percentile of the Fermi function.
The depth of this point below the gate, as determined from simulations, is subtracted from the cathode-gate distance to obtain the drift length $\Delta z$.
To measure the drift time from this point, we use \krthree{} calibration events.
Analogously to the simulated data, a Fermi function is fit to the ratio between the areas of the combined S2 signal and the first S1 signal, and its 50th percentile is used as the drift time.
This is seen in figure~\ref{fig:gateposition}.
To evaluate the systematic uncertainty on the drift velocity imposed by the arbitrary choice of the top of the drift region, we repeat the procedure for determining $\Delta t_\text{gate}$ and $\Delta z$ using the 16th and 84th percentiles of the Fermi function and use the difference as the systematic uncertainty.

The drift time from the cathode can be found using the fact that the ionisation electrons of events below it will drift downwards, and not result in an S2.
Events from the cathode will therefore have the largest observed drift time.
We also model the decrease of event rate with drift time near the cathode with a Fermi function, shown in figure~\ref{fig:cathodeposition}.
The cathode position is taken as the ($50 \pm 25$)th percentile, with the large uncertainty used in the absence of a well-motivated choice for the position within the distribution.
In both cases, we account for the systematic uncertainty due to the choice of percentile as well as the subdominant statistical uncertainty from the fit.

Because external fields leak into the drift region, the drift field is not exactly given by the potential difference between the cathode and gate electrode divided by their separation.
We use COMSOL to simulate the TPC in two dimensions, assuming axial symmetry, and with this determine the field throughout the drift region.
The resulting field strengths in the entire sensitive region (for the hermetic TPC) or with radius up to 23~mm (for the baseline TPC) are then histogrammed.
We present our results using the median of this histogram.
Uncertainties on the drift field are given by its 16th--84th percentile spread, to account for the non-uniform field causing the drift velocity to vary.
The hypothetical uniform field which would result in the measured drift velocity lies somewhere within the range of fields for each measurement.

\begin{figure}
    \centering
    \includegraphics{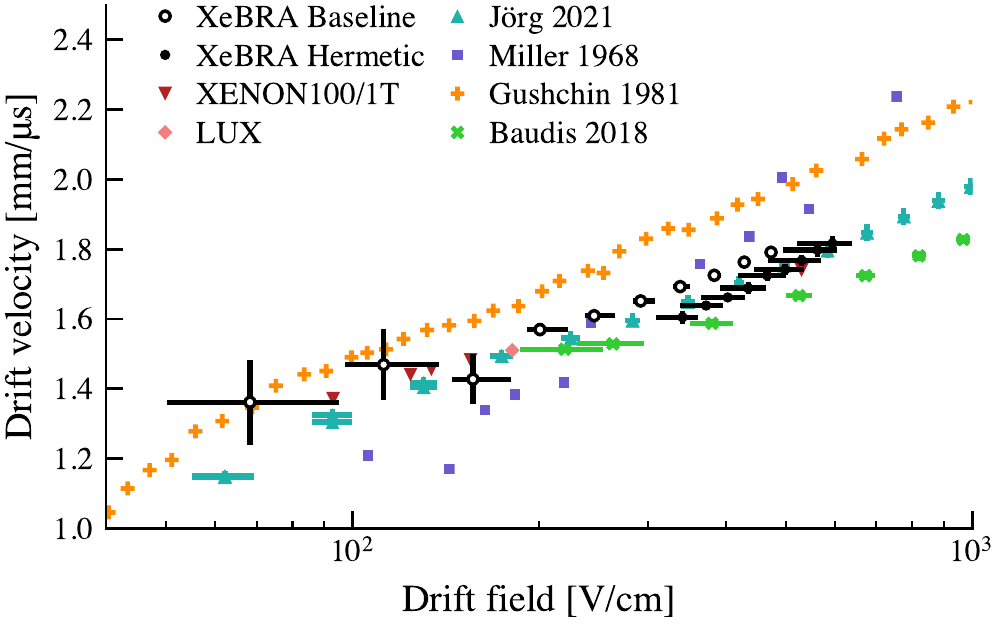}
    \caption{Field-dependence of the electron drift velocity in liquid xenon. We compare our results (hollow and filled black circles) to other recent publications~\cite{XurichTPC, MillerDriftSpeed, GushchinDriftSpeed, HeXeResults, XENON100Expt, XENONExpt, LUXFirstResults}.} \label{fig:driftvelocity}
\end{figure}

Figure~\ref{fig:driftvelocity} shows our results for the drift velocity as a function of the drift field.
We compare them to results from recent dark matter experiments~\cite{XENON100Expt, XENONExpt, LUXFirstResults} as well as dedicated measurements~\cite{XurichTPC, MillerDriftSpeed, GushchinDriftSpeed, HeXeResults}.
Our results are consistent with the general trend of the other measurements.
However, there is tension between different experiments, which could be attributable to varying operating temperatures~\cite{HeXeResults}, varying definitions of the S2 timing, as well as different methods being used to assess the drift field and maximum drift time.

Our results are also slightly inconsistent between our two detectors.
Since the analysis procedure was almost identical for both sets of results, this discrepancy is likely due to a physical difference which could be a different temperature within the sensitive region.
Another possible explanation is the lack of fiducialisation in the hermetic TPC.
This means that field inhomogeniety near the wall may affect the speed and direction of electron drift more than in the baseline TPC, where a 23~mm cut on radius is used.
That the drift field is less homogeneous in the hermetic TPC can be seen in the horizontal error bars of figure~\ref{fig:driftvelocity}.

\subsection{Charge and light yields}
\label{sec:yields}
The way the energy deposited by an interaction in liquid xenon is split between light and charge signals is important to distinguish nuclear recoil signals from electronic recoil backgrounds~\cite{AprileNRDiscrimination}.
Accordingly, it has been widely studied to provide inputs for future detector designs~\hbox{\citep[e.g][]{NEST, XurichTPC}}.
The splitting can be described in terms of the absolute light yield $L(E)$ and absolute charge yield $Q(E)$ which are the number of electrons and photons produced per unit energy deposited in an interaction.
We cannot directly measure the absolute yields, but the effective yields $L^\text{eff}(E)$ and $Q^\text{eff}(E)$, which are the mean corrected light and charge signals per unit energy:
\begin{align*}
    L^\text{eff}(E) &= \mathrm{cS1} / E; & Q^\text{eff}(E) &= \mathrm{cS2} / E.
\end{align*}
These yields are detector-specific, depending on the efficiency for detecting photons and on the S2~gain, i.e. the number of photons produced per electron.
To avoid this detector-dependence, we derive the relative charge and light yields:
\begin{align*}
    L^\text{rel}(E) &= L^\text{eff}(E)/L_0; & Q^\text{rel}(E) &= Q^\text{eff}(E)/Q_0.
\end{align*}
They are defined as fractions of the (hypothetical) light-only and charge-only yields $L_0$ and $Q_0$, which would be the yields if only one type of signal were produced.
The relative yields are universal across all detectors, since detection efficiencies appear in both $L^\text{eff}(E)$ and $L_0$ and therefore cancel.

\begin{figure}
    \includegraphics{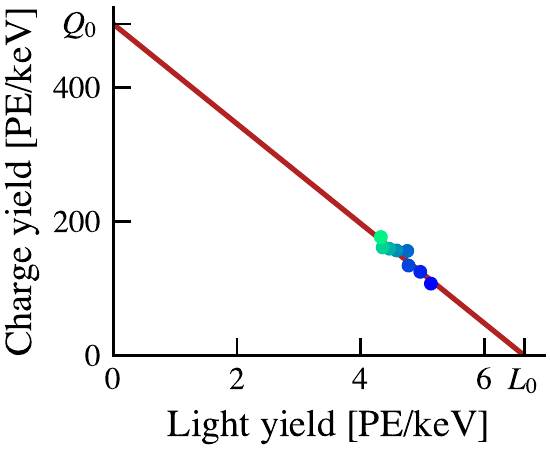}
    \hfill
    \includegraphics{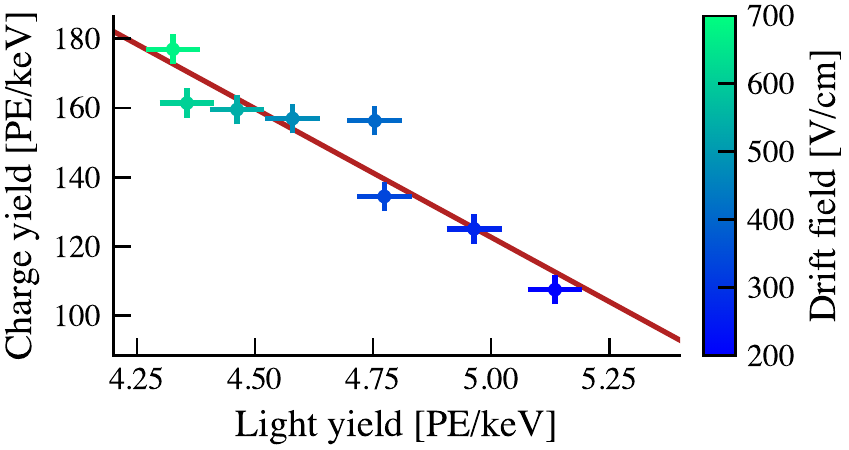}
    \centering
    \caption{Anticorrelation of light and charge signals from the 41.5~keV deposited by unresolved \krthree{} decays in the hermetic TPC at drift fields between 203~V/cm and 677~V/cm. Each point corresponds to one drift field, as indicated by the colour of the marker. The extraction field was kept at a constant 3.9~kV/cm. The left-hand plot shows the same data with the extrapolation to $L_0$ on the x-axis and $Q_0$ on the y-axis.}
    \label{fig:doke}
\end{figure}

We make use of the anticorrelation of the light and charge signals to determine $L_0$ and $Q_0$.
Assuming that the fraction of energy lost to heat is constant, the light and charge yields for all interactions fall on a single straight line when plotted against each other.
This is shown in figure~\ref{fig:doke} for the merged 41.5~keV signal from \krthree{} decays for various drift fields.
By extrapolating the best-fit to the x- and y-axes, we determine the light-only yield in the hermetic TPC to be $L_0 = (6.5 \pm 0.3)$~PE/keV and the charge-only yield to be $Q_0 = (520 \pm 50)$~PE/keV.

\begin{figure}
    \includegraphics{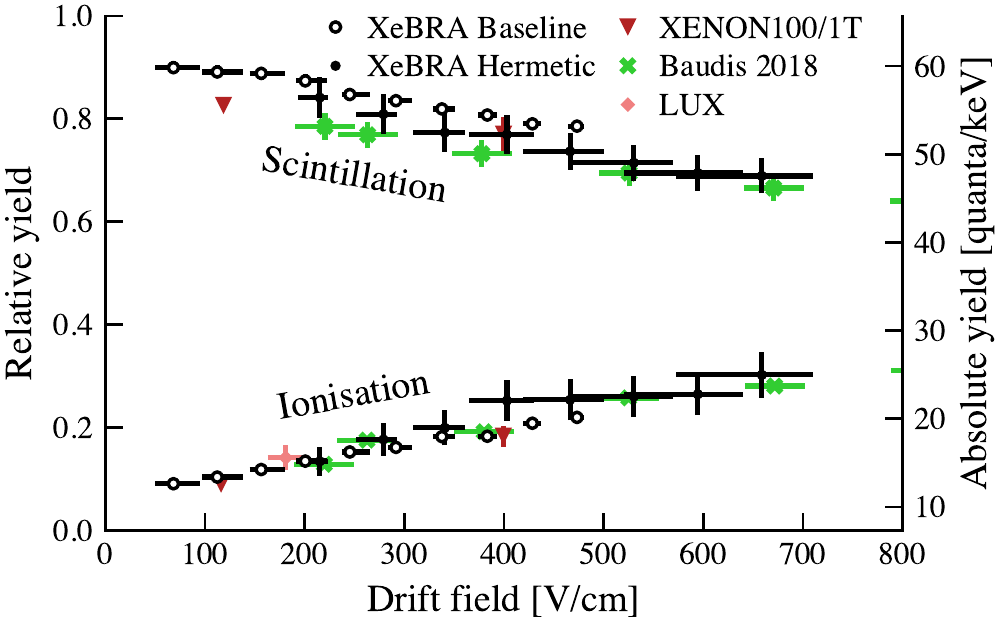}
    \centering
    \caption{%
        Relative light yield (top half) and relative charge yield (lower half) of the combined 41.5~keV \krthree{} deposition, depending on the drift field.
        These are expressed as a fraction of the hypothetical yields when only one of the two types of quantum is produced.
        The right axis shows the equivalent absolute yields, assuming $W=13.7~\1{eV}$.
        We compare our results (hollow and filled black circles) to those from dark matter experiments~\cite{Xe100Yields, XENONExpt, LUXReanalysis} and dedicated measurements in~\cite{XurichTPC}.}
    \label{fig:yields}
\end{figure}

The relative yields can then be computed as a function of drift field by taking the ratio of the observed yields to the light-only and charge-only yields.
These are shown in figure~\ref{fig:yields} and compared to results from other experiments.
Our results agree with those of large-scale dark-matter experiments and dedicated research detectors.
It is possible to convert the relative charge and light yields into absolute yields by making the approximation that on average it takes the same energy $W$ to create a single photon as to release a single electron.
We assume the commonly-used value $W = 13.7$~eV~\cite{DahlThesis}, although some more recent works suggest it might be smaller~\cite{EXOW,XurichW}.
The resulting absolute yields are shown on the second y-axis of figure~\ref{fig:yields}.

Using the ratio of our observed relative yields to the absolute yields, we also determine the photon-detection efficiency to be $g_1 = (0.122 \pm 0.002)$~PE/photon for the baseline TPC and $g_1 = (0.089 \pm 0.004)$~PE/photon for the hermetic TPC.
Similarly, the single-electron gain, or mean number of photoelectrons detected per released electron, is $g_2 = (5.49 \pm 0.05)$~PE/electron in the baseline TPC, with an extraction field (in the liquid xenon) of 2.8~kV/cm.
In the hermetic TPC, at an extraction field of 3.9~kV/cm, it is $g_2 = (7.1 \pm 0.7)$~PE/electron.

\section{Summary}

XeBRA is a platform for testing new liquid xenon TPC technologies at small scales and for measuring basic properties of the liquid xenon detector medium.
It has already been used to demonstrate the performance of a hermetic TPC, designed to reduce the radon level in the sensitive region~\cite{HermeticTPC}.
We have given an overview of the platform and its subsystems and described a baseline TPC, whose performance will be compared against a modified version of the TPC to investigate the viability of a single-phase, liquid-only TPC.
In such a TPC, electroluminescence within the liquid xenon produces the S2 signal. Previous publications have demonstrated the concept, albeit not in a TPC~\cite{ColumbiaSinglePhase}, and discussed its possible implications~\cite{SinglePhaseSim}.

We have also provided measurements of the electron drift velocity as a function of drift field.
Our results, which are compatible with the range of existing values, contribute to the pool of data available for the community.
However, at the moment the various measurements of the drift velocity are inconsistent within their uncertainties.
This might hint that there is a systematic difference between the results obtained with different detectors.
Resolving this discrepancy will require a future study investigating possible systematic effects as well as a consistent evaluation of the drift field and maximum drift time.

Furthermore, we determine the relative scintillation and ionisation yield for 41.5~keV energy depositions from a \krthree{} source at a range of fields.
Again, our measurements are compatible with previous results and provide useful input for simulations, which in turn are important tools to guide detector development.

\acknowledgments

This work is supported by the European Research Council (ERC) grant No.~724320 (ULTIMATE). We thank the members of the mechanical workshops of the Physikalisches Institut at the University of Freiburg and of the Laboratory for High Energy Physics at the University of Bern. We also thank all the bachelor students who wrote their theses on topics related to XeBRA.

\providecommand{\href}[2]{#2}\begingroup\raggedright\endgroup

\end{document}